\documentclass[10pt]{article}
\usepackage{amsmath}
\usepackage{amsfonts}
\usepackage{amssymb}
\usepackage{paralist}
\usepackage{graphics} 
\usepackage{epsfig} 
\usepackage[colorlinks=true]{hyperref}

\hypersetup{urlcolor=blue, citecolor=red}

\def\sqr#1#2{{\vcenter{\vbox{\hrule height.#2pt
              \hbox{\vrule width.#2pt height#1pt \kern#1pt \vrule
width.#2pt}
              \hrule height.#2pt}}}}
\def\3n{\negthinspace \negthinspace \negthinspace }
\def\2n{\negthinspace \negthinspace }
\def\1n{\negthinspace }

\def\L{\Lambda}

\def\Re{{\mathop{\rm Re}\,}}

\def\({\Big (}
\def\){\Big )}
\def\[{\Big[}
\def\]{\Big]}

\makeatletter
   
   \@addtoreset{equation}{section}
\makeatother

\def\ba{\begin{array}}
\def\be{\begin{equation}}
\def\bc{\begin{corollary}}
\def\bd{\begin{definition}}
\def\bea{\begin{eqnarray}}
\def\bel{\begin{equation}\label}
\def\bl{\begin{lemma}}
\def\bp{\begin{proposition}}
\def\br{\begin{remark}}
\def\bt{\begin{theorem}}
\def\ee{\end{equation}}
\def\eea{\end{eqnarray}}
\def\et{\end{theorem}}
\def\ec{\end{corollary}}
\def\el{\end{lemma}}
\def\ep{\end{proposition}}
\def\er{\end{remark}}
\def\ea{\end{array}}
\def\ed{\end{definition}}
\newtheorem{theorem}{Theorem}[section]

\newtheorem{corollary}{Corollary}[section]

\newtheorem{definition}{Definition}[section]

\newtheorem{lemma}{Lemma}[section]

\newtheorem{proposition}{Proposition}[section]
\newtheorem{remark}{Remark}[section]

\newtheorem{thm}{Theorem}[section]
\newtheorem{lem}{Lemma}[section]

\newtheorem{defi}{Definition}[section]

\title
      {Continuous-time Mean-Variance Portfolio Selection with Stochastic Parameters}

\author{Wan-Kai Pang\thanks{Wan-Kai Pang, Xun Li, and Ka-Fai Cedric Yiu are with the Department of Applied Mathematics, The Hong Kong Polytechnic University, Hunghom, Kowloon, Hong Kong},~~
Yuan-Hua Ni\thanks{ {Yuan-Hua Ni is with Department of Mathematics, Tianjin Polytechnic University, Tianjin 300160, PR China}
},~~ Xun Li, ~~Ka-Fai Cedric Yiu}

\begin{document}
\maketitle

%
%
%

\begin{abstract}

This paper studies a continuous-time market {under stochastic environment} where an agent, having
specified an investment horizon and a target terminal mean return,
seeks to minimize the variance of the return with
multiple stocks and a bond. In the considered model firstly proposed by \cite{Bielecki-Pliska}, the mean returns of individual assets are explicitly affected by
underlying Gaussian economic factors. Using past and present
information of the asset prices, a partial-information stochastic optimal control
problem with random coefficients is formulated. Here, the partial information is due to the fact that the economic factors can not be directly observed. Via dynamic programming
theory, the optimal portfolio strategy can be constructed by solving a deterministic forward Riccati-type ordinary differential equation and two linear deterministic backward ordinary differential equations.

\end{abstract}


\section{Introduction}

Mean-variance is by far an important investment
decision rule in financial portfolio selection, which is first proposed and
solved in the single-period setting by Markowitz in his Nobel-Prize-winning works \cite{Markowitz-1,Markowitz-2}. In these seminal papers, the variance of the final
wealth is used as a measure of the risk associated with the
portfolio and the agent seeks to minimize the risk of his investment
subject to a given mean return.
This model becomes the foundation of modern finance theory and inspires hundreds of extension and applications. For example, this leads to the elegant capital asset pricing model naturally \cite{SHARPE}.

The dynamic extension of the Markowitz model has been established in the subsequent years after the appearance of \cite{Markowitz-1,Markowitz-2}, employing heavily among others the martingale theory, convex duality and stochastic control. The pioneer work for continuous time case about multi-period portfolio management is \cite{Merton}. In \cite{Merton}, Merton uses dynamic programming and partial differential equation (PDE) theory to derive
and analyze the relevant Hamilton-Jacobi-Bellman (HJB) equation, and thus obtains the optimal strategy. Alternatively, to avoid dynamic programming, in \cite{Pliska}, the author introduce the so-called risk-neutral (martingale)
probability measure in order to reduce the computational difficulties associated with PDEs. In \cite{Zhou-li}, the authors formulate the mean-variance problem with deterministic coefficients to a linear-quadratic (LQ) optimal problem. As there is no running cost in the objective function, this formulation is inherently an indefinite stochastic LQ control problem. As extensions of \cite{Zhou-li}, for example, \cite{Lim-zhou} deals with random coefficients case; while \cite{zhou-yin} considers regime switching market. For discrete time case, \cite{Li-Ng} completely solves the multiperiod mean-variance portfolio selection problem. Analytical optimal strategy and an efficient algorithm to find this strategy are proposed. For more about the history of the mean-variance model, \cite{Steinbach} and \cite{Bielecki} are refered.

In \cite{Bielecki-Pliska}, in order to tackle the  the computational
tractability and the statistical difficulties associated with the estimation of model
parameters, Bielecki and Pliska introduce a model
such that the underlying economic factors such as accounting ratios, dividend yields, and macroeconomic measures are
explicitly incorporated in the model. Exactly, the factors are assumed to follow Gaussian processes and
the drifts of the stocks are linear functions of these factors. This model motivates many subsequent researches; see, for example, \cite{Nagai-Peng} and \cite{Bielecki-Pliska-2}.
{In practice, investors can only observe past and present
asset prices to decide his current portfolio strategy;
and, random factors cannot be completely observed. Therefore, the underlying
problem falls into the category of portfolio selection under partial
information}. A significant progress in the realm of mean-variance
concerning partial information is the work of \cite{Xiong-Zhou}. In \cite{Xiong-Zhou}, a separation principle is shown to hold in this partial information setting; efficient strategies based on the partial information are derived, which involve the optimal filter of the stock appreciation rate processes; in addition, the particle system representation of the obtained filter is employed to develop analytical and numerical approaches. It is valuable to point out that backward stochastic differential equations (BSDEs) methodology is employed to tackle this problem.

This paper attempts to deal with the mean-variance portfolio selection
under partial information based on the model of \cite{Bielecki-Pliska}. By exploiting the properties of the filter process and the wealth process, we tackle this problem directly by the dynamic programming
approach. We show that optimal strategy can be constructed by solving a \emph{deterministic} forward Riccati-type ordinary differential equation (ODE) and a system of linear \emph{deterministic} backward ODEs. Clearly, by reversing the time, a deterministic backward ODE can be converted to a forward one. Therefore, we can easily derive the analytic solutions of the ODEs, and thus the analytic form of the optimal strategies. This is the main contribution of the note. The proposed procedure is different from that of \cite{Xiong-Zhou}, where BSDEs are employed.

The rest of the paper is organized as follows. In Section 2, we
formulate the mean-variance portfolio selection model under partial
information, and an auxiliary problem is introduced. Section 3 gives the optimal strategy of the auxiliary problem by dynamic programming method. Section 4 studies the original
problem, while Section 5 gives some concluding remarks.


\section{Mean-Variance Model}


Throughout this paper
$(\Omega,\mathcal{F},P,\{\mathcal{F}_t\}_{t\geq0})$ is a fixed
filtered complete probability space on which defined a standard
$\mathcal{F}_t$-adapted $(n+m)$-dimensional Brownian motion $\{W(t),
t\geq 0\}$ with $W(t)\equiv(W^1(t),\cdots,W^{n+m}(t))^T$ and
$W(0)=0$. Let $T>0$ be the terminal time of an investment,
and $L^2_\mathcal{F}(0,T;\mathbb{R}^d)$ denote the set of all
$\mathbb{R}^d$-valued, $\mathcal{F}_t$-adapted stochastic processes $f(t)$ with $E\int_0^T|f(t)|^2dt<+\infty$,
similarly $L^2_{\mathcal{H}}(0,T;\mathbb{R}^l)$ can be defined for
any functions with domain in $\mathbb{R}^l$ and filtration
$\mathcal{H}_t$.

There is a capital market containing $m+1$ {basic securities}
(or assets) and $n$ economic factors. The securities consist
of a bond and $m$ stocks. The set of factors may include short-term
interests, the rate of inflation, and other economic factors \cite{Bielecki-Pliska-2}. One of the securities is a risk-free bank account whose {value
process} $S_0(t)$ is subject to the following ordinary differential
equation
\begin{equation}\label{equ-r}
dS_0(t)= r(t)S_0(t)dt,\;\; t\geq0,~~
S_0(0) = s_0>0,
\end{equation}
where $r(t)$ is the interest rate, a deterministic function of $t$.
The other $m$ assets are risky {stocks} whose price processes
$S_1(t),\cdots, S_m(t)$ satisfy the following stochastic
differential equations (SDEs)
\begin{equation}\label{equ-stock}
\left\{\begin{array}{rcl}
dS_i(t) & = & S_i(t)\big\{\mu_i(t)dt+\sum_{j=1}^{n+m}\sigma_{ij}(t) dW^j(t)\big\},\;\; t\geq0,\\
 S_i(0) & = & s_i>0,\;\; i=1,2,\cdots,m,
\end{array}\right.
\end{equation}
where $\mu_i(t),i=1,...,m$, are the {appreciation rates}, and
$\sigma_i(t),i=1,...,m$ are the deterministic {volatility} or
dispersion rate of the stocks. In this paper, we assume that the
appreciation rates are affine functions of the mentioned economic factors,
and the factors are Gaussian processes. To be precise, denoting
$(\mu_1(t),\mu_2(t),...,\mu_m(t))^T$ by $\mu(t)$, we have
\begin{eqnarray*}\label{equ-V}
\left\{\begin{array}{rcl}\mu(t)&=&a+Ay(t)\triangleq \mu^y(t), ~a\in \mathbb{R}^m,~A\in \mathbb{R}^{m\times n},\\
dy(t)&=&(d+Dy(t))dt+\Lambda dW(t),~y(0)=y_0\in \mathbb{R}^n,
\end{array}\right.\end{eqnarray*}
where the constant matrices $d,D,\Lambda$ are of $n\times1$, $n\times
n,n\times (m+n)$, respectively.

Consider an agent with an initial endowment $x_0>0$ and an
investment horizon $[0,T]$, whose total wealth at time $t\in[0,T]$
is denoted by $X(t)$. Assuming that the trading of shares is
self-financed and takes place continuously, and that transaction
cost and consumptions are not considered, then $X(t)$ satisfies
(see, e.g., \cite{Karatzas-Shreve})
\begin{equation}\label{systems}
\left\{\begin{array}{rcl} dX(t) & = & \Big\{r(t)X(t) +
\sum_{i=1}^m\big[\mu_i^y(t)-r(t)\big]\pi_i(t)\Big\}dt
 +\sum_{i=1}^m\sum_{j=1}^{n+m}\pi_i(t)\sigma_{ij}(t)dW^j(t), \\ 
 X(0) & = & x_0,
\end{array}\right.
\end{equation}
where $\pi_i(t),\;\; i=1,2\cdots,m,$ denote the total market value
of the agent's wealth in the $i$-th stock. We call the process
$\pi(t) = (\pi_1(t),\cdots,\pi_m(t))^T$, $0\leq t\leq T$, a {\it
portfolio} of the agent.

Let
\begin{eqnarray*}
&&S(t)=(S_1(t),...,S_m(t))^T,\\
&&\mathcal{G}_t=\sigma(S_0(u),S(u): ~u\leq t),t\geq 0.
\end{eqnarray*}
As pointed out by \cite{Xiong-Zhou}, practically, the investor
can only observe the prices of assets. So, at time $t$, the information
that available to the investor is the past and present assets'
prices, equivalently, the filtration $\mathcal{G}_t$. Thus, the
investor's strategy should be based on his/her available
information. Therefore, $\pi_t$ should be
$\mathcal{G}_t$-measurable. To be exact, we
define the following admissible portfolio.

\begin{defi}\label{def001}
{\rm  A portfolio $\pi(\cdot)$ is said to be {\it admissible}
if $\pi(\cdot) \in {L}_\mathcal{G}^2(0,T;\Re^m)$ 
and the SDE (\ref{systems}) has a unique solution $x(\cdot)$
corresponding to $\pi(\cdot)$. The totality of all admissible
portfolios is denoted by $\Pi$. 
}
\end{defi}

The agent's objective is to find an admissible portfolio
$\pi(\cdot)$, among all such admissible portfolios that his/her
expected terminal wealth $EX(T) = \bar{x}$, where $\bar{x}\geq x_0
e^{\int_0^T r(t)dt}$ is given {\it a priori}, so that the risk
measured by the variance of the terminal wealth
\begin{equation}\label{var0}
\mbox{\rm Var}\; X(T) := E[X(T) - EX(T)]^2 \equiv E[X(T) -
\bar{x}]^2
\end{equation}
is minimized. The problem of finding such a portfolio $\pi(\cdot)$
is referred to as the {\it mean-variance portfolio selection
problem}. Mathematically, we have the following formulation.

\begin{defi}\label{def002}
{\rm The mean--variance portfolio selection problem, with respect to
the initial wealth $x_0$, is formulated as a constrained stochastic
optimization problem parameterized by $\bar{x}\geq x_0  e^{\int_0^T
r(t)dt}$:
\begin{equation}\label{equ007}
\left\{\begin{array}{ll}
\mbox{minimize} &Var X(T)= E[X(T) - \bar{x}]^2 = E[X(T)^2] - \bar{x}^2, \\
\mbox{subject to} & \left\{
\begin{array}{l}
X(0) = x_0,\;\;EX(T) = \bar{x}, \\
(X(\cdot),\pi(\cdot)) \; \mbox{ admissible}.
\end{array}\right.
\end{array}\right.
\end{equation}
The problem is called {\it feasible} (with respect to $\bar{x}$) if
there is at least one admissible portfolio satisfying $EX(T) =
\bar{x}$. An optimal portfolio, if it ever exists, is called an {\it
efficient portfolio strategy} with respect to $\bar{x}$, and $Var
X(T)$ is called an efficient point. The set of all efficient
points is obtained when the parameter $\bar{x}$ varies between
$[x_0e^{\int_0^T r(s)ds},+\infty )$ }.
\end{defi}




We impose the basic assumptions of this paper.

\emph{Assumption (PD)}. For any  $t\geq 0$, $\sigma(t)\sigma^T(t)>0$.

\begin{remark}
This assumption is popular in the literatures about portfolio selection; see, for example, \cite{Bielecki-Pliska}, \cite{Bielecki-Pliska-2}, \cite{Li-Zhou}, \cite{Nagai-Peng}, \cite{Xiong-Zhou}.
\end{remark}

Let
\begin{eqnarray*}\label{equ-B}
B(t)\triangleq \left(\mu^y(t)\right)^T-r(t)\mathbf{1},~~~
\sigma(t)=(\sigma_{ij}(t))_{m\times(n+m)},
\end{eqnarray*}
with $\mathbf{1}$ being a $m$-dimensional row vector with all its
entries being 1. Then,
(\ref{systems}) can be rewritten as
\begin{equation}\label{system2}
\left\{\begin{array}{l}
dX(t) = [r(t)X(t)+B(t)\pi(t)]dt+\pi(t)'\sigma(t) dW(t), \\ 
X(0) = x_0.
\end{array}\right.
\end{equation}
By the definition of $\pi$, our problem falls into the category of
stochastic control based on partial information. Here, the partial
information means that we cannot know the process $y(t)$, and thus
 $B(t)$. In order to design admissible strategy, we firstly
need to derive the optimal estimation of $y_t$. Let
\begin{eqnarray*}&&\Gamma=\sigma(t)\sigma^T(t)\in \mathbb{R}^{m\times m},\\
&&\overline{\Gamma}=(\Gamma_{11}(t),...,\Gamma_{nn}(t))^T,~~\Sigma(t)=\Gamma^{\frac{1}{2}}(t)=\Sigma^T(t),\\
&&Y(t)=(\log S_1(t),...,\log S_m(t))^T\triangleq \log S(t).
\end{eqnarray*}
By It\^o 's formula we have
\begin{eqnarray*}
dY(t)=\left(a+Ay(t)-\frac{1}{2}\overline{\Gamma}(t)\right)dt+\sigma(t)
dW_t,\quad Y_0=\log S(0).
\end{eqnarray*}
Define
\begin{eqnarray}\label{v}
dv(t)=\Sigma^{-1}\left[dY(t)-\left(a+A\hat{y}(t)-\frac{1}{2}\overline{\Gamma}(t)\right)dt\right],
\end{eqnarray}
then $\{v(t),t\geq 0\}$ is a Brownian motion under the original
probability measure (Liptser and Shiryaev (2001)). The estimation of
$y(t)$ is given by (Theorem 10.3 of \cite{Shiryaev})
\begin{eqnarray}\label{filtering}
\left\{\begin{array}{l}d\hat{y}(t)=(d +
D\hat{y}(t))dt+\left(\Lambda\sigma^T+\beta(t)
A^T \right)(\Sigma^T)^{-1}dv(t),\\
\dot{\beta}(t)=D\beta(t)+\beta(t)D^T+\Lambda\Lambda^T-(\Lambda\sigma^T+\beta(t)A^T)
(\Sigma\Sigma^T)^{-1}(\Lambda\sigma^T+\beta(t)A^T)^T,\\
\hat{y}(0)=y_0,~\beta(0)=0.
\end{array}\right.\end{eqnarray}
By (\ref{v}), a simple calculation shows that
\begin{eqnarray}\label{v-2}
\sigma(t)dW(t)=\Sigma(t)dv(t)+A[\hat{y}(t)-y(t)].
\end{eqnarray}
Substituting (\ref{v-2}), we have an equivalent representation of
the wealth process
\begin{eqnarray}\label{separation}
\left\{\begin{array}{l}dX(t) =
[r(t)X(t)+\overline{B}(t)\pi(t)]dt+\pi(t)'\Sigma(t)
dv(t),\\
X(0)=x_0,\end{array}\right.
\end{eqnarray}
where
\begin{eqnarray}\label{B-new}
\overline{B}(t)=(a+A\hat{y}(t))^T-r(t)\mathbf{1}.
\end{eqnarray}
This is the separation principle developed by \cite{Xiong-Zhou},
which enables us to solve problem (\ref{equ007}) as if the
appreciation rate $\mu^y(\cdot)$ were known, and then replace
$\mu^y(\cdot)$ by its optimal estimation. So, (\ref{equ007}) can be
equivalently formulated as
\begin{equation}\label{new-formulation} \left\{\begin{array}{ll}
\mbox{minimize} & E[(X(T) - \bar{x})^2], \\
\mbox{subject to}&\left\{\begin{array}{l}EX(T)=\bar{x},\\\pi\in \Pi,\\
(X(\cdot),\pi(\cdot))\mbox{ satisfy
}(\ref{filtering})(\ref{separation})(\ref{B-new}).
\end{array}\right.\end{array}\right.
\end{equation}
By general convex optimization theory, the constrained optimal
problem (\ref{new-formulation}) with ($EX(T)=\bar{x}$)  can be converted
into an unconstrained one by introducing a Lagrange multiplier
$\gamma$. To be concrete, for any fixed $\gamma$, we consider the
following problem
\begin{equation}\label{equ008}
\left\{\begin{array}{ll}
\mbox{minimize} & E[X(T)-\bar{x}]^2 - 2\gamma E[X(T)-\bar{x}] = E[X(T) - \gamma-\bar{x}]^2 - \gamma^2, \\
\mbox{subject to} & \left\{\begin{array}{l}\pi\in
\Pi,\\(X(\cdot),\pi(\cdot))\mbox{ satisfy
}(\ref{filtering})(\ref{separation})(\ref{B-new}),
\end{array}\right.
\end{array}\right.
\end{equation}
which is equivalent to the following (denoting
$\bar{x}+\gamma$ by $\alpha$ for any fixed $\gamma$)
\begin{equation}\label{equ009} \left\{\begin{array}{ll}
\mbox{minimize} & \frac{1}{2}E[(X(T) - \alpha)^2], \\
\mbox{subject to} & \left\{\begin{array}{l}\pi\in
\Pi,\\(X(\cdot),\pi(\cdot))\mbox{ satisfy
}(\ref{filtering})(\ref{separation})(\ref{B-new}),
\end{array}\right.
\end{array}\right.
\end{equation}
in the sense that two problems have exactly the same optimal
strategy.
In the following, we will call problem (\ref{equ009}) {the
auxiliary problem} of the original problem (\ref{new-formulation}).

\section{Optimal Policy for the Auxiliary Problem}

The problem (\ref{equ009}) can be viewed as an unconstrained special
stochastic optimal control problem with random coefficients in
system equation and zero integral term in the performance index.
Different from existing results using BSDEs methodology, in this section, we intend to
derive the optimal portfolio strategy from dynamic programming
directly. This enables us to derive the optimal policy by
solving just two linear deterministic backward ODEs and a
Riccati-type forward deterministic ODE.

\subsection{Analysis of Hamilton-Jacobi-Bellman equation}

Let $J(t,X,\hat{y})$ denote the performance of problem
(\ref{equ009}) at time $t$, with boundary condition
$J(T,X,\hat{y})=\frac{1}{2}E[(X(T) -\alpha)^2] $. Then, it is
evident that the following HJB equation
is satisfied
\begin{eqnarray}\label{HJB}
\min_{\pi\in\Pi}(\mathcal{L}J)(t,X,\hat{y})=0,~~J(T,X,\hat{y})=\frac{1}{2}(X-\alpha)^2,
\end{eqnarray}
where $\mathcal{L}$ is the infinitesimal generator operator of the
closed system (\ref{filtering})(\ref{separation})(\ref{B-new}), and
the independence of $X$ on policy $\pi$ is suppressed.

To evaluate $\mathcal{L}$, first of all, by
(\ref{filtering})(\ref{separation}) we have
\begin{eqnarray*}\label{covriance }
dX(t)d\hat{y}(t)&=&\pi^T\Sigma
dv(t)\cdot\left(\Lambda\sigma^T+\beta(t) A^T \right)(\Sigma^T)^{-1}dv(t)\nonumber\\
&=&\left(\Lambda\sigma^T+\beta(t) A^T
\right)(\Sigma^T)^{-1}dv(t)dv(t)^T\Sigma^T\pi\nonumber\\
&=&\left(\Lambda\sigma^T+\beta(t) A^T \right)\pi dt.
\end{eqnarray*}
By It\^o's formula, it follows that
\begin{eqnarray*}
\mathcal{L}J&=&J_t+J_X(rX+\overline{B}\pi)+J_{\hat{y}}^T
(d+D\hat{y})+\frac{1}{2}J_{XX}\pi^T \Sigma\Sigma^T\pi+J_{X\hat{y}}^T
\left(\Lambda\sigma^T+\beta A^T \right)\pi\\
&&+\frac{1}{2}Tr\left[ \left(\Lambda\sigma^T+\beta A^T
\right)(\Sigma\Sigma^T)^{-1}\left(\Lambda\sigma^T+\beta A^T
\right)^T J_{\hat{y}\hat{y}} \right],
\end{eqnarray*}
where $J_t$ is the partial derivative of $J$ with respect to $t$,
$J_{XX}$ is the second order partial derivative of $J$ with respect to
$X$, and $J_{X},J_{X\hat{y}},J_{\hat{y}\hat{y}}$ are similarly
defined.
On the assumption that $J_{XX} > 0$, we get the following optimal strategy
\begin{eqnarray}\label{optimal control-1}
\pi=-(\Sigma\Sigma^T)^{-1}\left[\overline{B}^T\frac{J_X}{J_{XX}}+(\Lambda\sigma^T+\beta
A^T)^T \frac{J_{X\hat{y}}}{J_{XX}} \right],
\end{eqnarray}
which makes $\mathcal{L}J$ minimal.
Substituting (\ref{optimal control-1}) into (\ref{HJB}) leads to
\begin{eqnarray}\label{HJB-3}
\left\{\begin{array}{l} J_t+rXJ_X+J_{\hat{y}}^T(d+D\hat{y})+
\frac{1}{2}Tr\left[ \left(\Lambda\sigma^T+\beta A^T
\right)(\Sigma\Sigma^T)^{-1}\left(\Lambda\sigma^T+\beta A^T
\right)^T J_{\hat{y}\hat{y}} \right]\\
-\frac{1}{2}\left[\overline{B}^T\frac{J_X}{J_{XX}}+(\Lambda\sigma^T+\beta
A^T)^T \frac{J_{X\hat{y}}}{J_{XX}}
\right]^T(\Sigma\Sigma^T)^{-1}\left[\overline{B}^T\frac{J_X}{J_{XX}}+(\Lambda\sigma^T+\beta
A^T)^T
\frac{J_{X\hat{y}}}{J_{XX}} \right]{J_{XX}}=0,\\
J(T,X,\hat{y})=\frac{1}{2}(X-\beta)^2.
\end{array}\right.
\end{eqnarray}
In this and the following PDEs and ODEs, the arguments $t,X,\hat{y}$ are
always suppressed to simplify the notations.

Noticing that the terminal condition of $J$ is a nonhomogeneous
function of $X$, in order to make (\ref{HJB-3}) homogeneous, we
set
\begin{eqnarray}\label{z}
z(t)=X(t)-\alpha e^{-\int_t^Tr(s)ds}.
\end{eqnarray}
Simple calculation shows
\begin{eqnarray*}
&&J(t,X,\hat{y})=J(t,z+\alpha
e^{-\int_t^Tr(s)ds},\hat{y})\triangleq H(t,z,\hat{y}),\\
&&H_t=J_t-\alpha J_X r(t)e^{-\int_t^Tr(s)ds},\\
&&H_X=H_zz_X=H_z,~H_{XX}=H_{zz},~H_{X\hat{y}}=H_{z\hat{y}}.
\end{eqnarray*}
Substituting $z$ and the above equalities into (\ref{HJB-3}), we
obtain that
\begin{eqnarray}\label{HJB-0}
\left\{\begin{array}{l}
H_t+rzH_z+H_{\hat{y}}^T(d+D\hat{y})+\frac{1}{2}Tr\left[
\left(\Lambda\sigma^T+\beta A^T
\right)(\Sigma\Sigma^T)^{-1}\left(\Lambda\sigma^T+\beta A^T
\right)^T H_{\hat{y}\hat{y}} \right]\\
-\frac{1}{2}\left[\overline{B}^T\frac{H_z}{H_{zz}}+(\Lambda\sigma^T+\beta
A^T)^T \frac{H_{z\hat{y}}}{H_{zz}}
\right]^T(\Sigma\Sigma^T)^{-1}\left[\overline{B}^T\frac{H_z}{H_{zz}}+(\Lambda\sigma^T+\beta
A^T)^T\frac{H_{z\hat{y}}}{H_{zz}} \right]{H_{zz}}=0,\\
H(T,z,\hat{y})=\frac{1}{2}z^2.
\end{array}\right.
\end{eqnarray}

By the special structure of (\ref{HJB-0}), the following separation form of $H(t,z,\hat{y})$ is taken
\begin{eqnarray}\label{H}
H(t,z,\hat{y})=\frac{1}{2}f(t,\hat{y})z^2,~\mbox{with}~f(T,\hat{y})=1~~\mbox{for
all}~\hat{y},
\end{eqnarray}
whose reasonableness will be proved in Theorem 3.1.
Therefore, the optimal control (\ref{optimal control-1}) has the following
structure
\begin{eqnarray*}
\pi=-(\Sigma\Sigma^T)^{-1}\left[\overline{B}^T+(\Lambda\sigma^T+\beta
A^T)^T \frac{\partial \ln f}{\partial \hat{y}} \right]z ,
\end{eqnarray*}
which is linear in $z$, and (\ref{HJB-0}) is equivalent to
\begin{eqnarray}\label{HJB-4}
\left\{\begin{array}{l}\frac{1}{2}\frac{\partial f}{\partial
t}z^2+rfz^2+\frac{1}{2}\frac{\partial f}{\partial
\hat{y}}^T(d+D\hat{y})z^2 +\frac{1}{2}Tr\left[
\left(\Lambda\sigma^T+\beta A^T
\right)(\Sigma\Sigma^T)^{-1}\left(\Lambda\sigma^T+\beta A^T
\right)^T\frac{\partial^2f}{\partial\hat{y}^2}
\right]z^2\\
-\frac{1}{2}\left[\overline{B}^T z+(\Lambda\sigma^T+\beta A^T)^T
\frac{\partial \ln f}{\partial \hat{y}}z
\right]^T(\Sigma\Sigma^T)^{-1}\left[\overline{B}^T
z+(\Lambda\sigma^T+\beta A^T)^T \frac{\partial \ln f}{\partial
\hat{y}}z \right]f=0,\\\frac{1}{2}f(T,\hat{y})z^2=\frac{1}{2}z^2.
\end{array}\right.
\end{eqnarray}
Clearly, if $f(t,\hat{y})$ solves the following PDE
\begin{eqnarray}\label{HJB-5}
\left\{\begin{array}{l}\frac{\partial f}{\partial
t}+2rf+\frac{\partial f}{\partial \hat{y}}^T(d+D\hat{y})+Tr\left[
\left(\Lambda\sigma^T+\beta A^T
\right)(\Sigma\Sigma^T)^{-1}\left(\Lambda\sigma^T+\beta A^T
\right)^T\frac{\partial^2f}{\partial\hat{y}^2}
\right]\\
~~-\left[\overline{B}^T+(\Lambda\sigma^T+\beta A^T)^T \frac{\partial
\ln f}{\partial \hat{y}}
\right]^T(\Sigma\Sigma^T)^{-1}\left[\overline{B}^T
+(\Lambda\sigma^T+\beta A^T)^T \frac{\partial \ln
f}{\partial \hat{y}} \right]f=0,\\ f(T,\hat{y})=1,\\
\overline{B}(t)=(a+A\hat{y}(t))^T-r(t)\mathbf{1} ,
\end{array}\right.
\end{eqnarray}
then $H(t,z,\hat{y})$ has the explicit form of (\ref{H}).

\subsection{Optimal Policy}

Notice that the left hand side of the first equation in
(\ref{HJB-5}) is linear in $f, \frac{\partial f}{\partial
t}, \hat{y}, \frac{\partial^2
f}{\partial \hat{y}^2}$, and quadratic in $\frac{\partial\ln f}{\partial \hat{y}}$. Therefore, we assume that $f$ has the
following expression
\begin{eqnarray}\label{f}
f(t,\hat{y})=\exp\left\{p(t)+q^T(t)\hat{y}+\hat{y}^T G(t)\hat{y}
\right\},
\end{eqnarray}
with $p(t)\in \mathbb{R}, q(t)\in \mathbb{R}^n, G(t)\in S^{n\times
n}$ to be specified later. Here, $S^{n\times n}$ denotes the set
of all symmetric $n\times n$ real matrices.
The form (\ref{f}) of $f$  enables us to get an equivalent equation
that is independent of $f$ and is only a quadratic function of
$\hat{y}$. Fixing the coefficients of the obtained equation to be
zero, we can determine $p,q,G$ by solving several
equations. Thus, we may prove that $H$ given in (\ref{H}) satisfied the HJB equation (\ref{HJB}), indeed. Therefore, we have
the following theorem.
\begin{thm}\label{thm-1}
For problem (\ref{equ009}), the optimal strategy is given by
\begin{eqnarray}\label{optimal control-last one0-0-0}
\pi(t)&=&-(\Sigma(t)\Sigma^T(t))^{-1}\left[a^T-r(t)\mathbf{1}+(\Lambda\sigma^T(t)+\beta(t)A^T)^T
q(t)\right.\nonumber\\
&&~~\left.+\left(A+(\Lambda\sigma^T(t)+\beta(t)A^T)^T
G(t)\right)\hat{y}(t) \right]\left(X(t)-\alpha
e^{-\int_t^Tr(s)ds}\right),\end{eqnarray} where $\beta(t),q(t),G(t)$
are the unique solutions to the second equation of (\ref{filtering}) and following ODEs,
respectively,
\begin{eqnarray}\label{three-equation-2-0-0} &&\left\{\begin{array}{l}\frac{d q}{d
t}^T+\left[q^T D+ d^T
G\right]-2(a^T-r\mathbf{1})(\Sigma\Sigma^T)^{-1}A\\
~~-2(a^T-r\mathbf{1})(\Sigma\Sigma^T)^{-1}(\Lambda\sigma^T+\beta
A^T)^T
G-2q^T(\Lambda\sigma^T+\beta A^T)(\Sigma\Sigma^T)^{-1}A=0,\\
q(T)=0,\end{array}\right.\\
&&\label{three-equation-3-0-0} \left\{\begin{array}{l}\frac{d G}{d
t}-A^T
(\Sigma\Sigma^T)^{-1}A-\left[A^T(\Sigma\Sigma^T)^{-1}(\Lambda\sigma^T+\beta
A^T)^T+D^T\right]G\\
~~-G\left[(\Lambda\sigma^T+\beta
A^T)(\Sigma\Sigma^T)^{-1}A+D\right]=0,\\
G(T)=0.\end{array}\right.
\end{eqnarray}
\end{thm}

\emph{{Proof}}. Bearing the form (\ref{f}) of $f$ in mind,
simple calculation shows that
\begin{eqnarray*}
\frac{\partial f}{\partial \hat{y}}&=&f\left(q+G\hat{y} \right),\\
\frac{\partial^2f}{\partial \hat{y}^2}&=&f\left(qq^T+q\hat{y}^T
G+G\hat{y}q^T+G\hat{y}\hat{y}^T G+G \right),\\
\frac{\partial \ln f}{\partial\hat{y}}&=&q+G\hat{y},\\
\frac{\partial f}{\partial t}&=&f\left(\frac{d p(t)}{d t}+\frac{d
q(t)}{d t}\hat{y}+\hat{y}^T\frac{d G(t)}{d t}\hat{y}\right).
\end{eqnarray*}
Therefore, (\ref{HJB-5}) is equivalent to
\begin{eqnarray}\label{HJB-6}
\left\{\begin{array}{l}\frac{d p}{d t}+\frac{d q}{d
t}^T\hat{y}+\hat{y}^T\frac{d
G}{d t}\hat{y}+2r+(q+G\hat{y})^T(d+D\hat{y})\\
+Tr\left[ \left(\Lambda\sigma^T+\beta A^T
\right)(\Sigma\Sigma^T)^{-1}\left(\Lambda\sigma^T+\beta A^T
\right)^T (qq^T+q\hat{y}^T
G+G\hat{y}q^T+G\hat{y}\hat{y}^T G+G )\right]\\
-\left[\overline{B}^T+(\Lambda\sigma^T+\beta A^T)^T (q+G\hat{y})
\right]^T(\Sigma\Sigma^T)^{-1}\times\left[\overline{B}^T +(\Lambda\sigma^T+\beta
A^T)^T (q+G\hat{y}) \right]=0,\\
p(T)+q^T(T)\hat{y}(T)+\hat{y}^T(T) G(T)\hat{y}(T)=0,\end{array}\right.
\end{eqnarray}
which is equivalent to
\begin{eqnarray}\label{HJB-6}
\left\{\begin{array}{l}\frac{d p}{d t}+\frac{d q}{d
t}^T\hat{y}+\hat{y}^T\frac{d
G}{d t}\hat{y}+2r+(q+G\hat{y})^T(d+D\hat{y})\\
~~+Tr\left[ \left(\Lambda\sigma^T+\beta A^T
\right)(\Sigma\Sigma^T)^{-1}\left(\Lambda\sigma^T+\beta A^T
\right)^T G \right]\\
~~-[(a+A\hat{y})^T-r\mathbf{1}](\Sigma\Sigma^T)^{-1}[(a+A\hat{y})^T-r\mathbf{1}]^T\\
~~-2[(a+A\hat{y})^T-r\mathbf{1}](\Sigma\Sigma^T)^{-1}\left(\Lambda\sigma^T+\beta
A^T \right)^T(q+G\hat{y})=0,\\
p(T)+q^T(T)\hat{y}(T)+\hat{y}^T(T) G(T)\hat{y}(T)=0.
\end{array}\right.
\end{eqnarray}

The left hand of above PDE can be decomposed into the
three terms:

1). the term that is irrespective of $\hat{y}$
\begin{eqnarray*}
&&\frac{d p}{d t}+q^T\left(\Lambda\sigma^T+\beta A^T
\right)(\Sigma\Sigma^T)^{-1}\left(\Lambda\sigma^T+\beta A^T
\right)^T q\\
&&
-\left[2(a^T-r\mathbf{1})(\Sigma\Sigma^T)^{-1}(\Lambda\sigma^T+\beta
A^T)^T-d^T \right]q+2r
-(a^T-r\mathbf{1})(\Sigma\Sigma^T)^{-1}(a^T-r\mathbf{1})^T \\
&&+Tr\left[\left(\Lambda\sigma^T+\beta A^T
\right)(\Sigma\Sigma^T)^{-1}\left(\Lambda\sigma^T+\beta A^T
\right)^T G\right];
\end{eqnarray*}

2). the term that is linear in $\hat{y}$
\begin{eqnarray*}
&&\frac{d q}{d t}^T\hat{y}+\left[q^T D+ d^T
G\right]\hat{y}-2(a^T-r\mathbf{1})(\Sigma\Sigma^T)^{-1}A\hat{y}\\
&&-2(a^T-r\mathbf{1})(\Sigma^2)^{-1}(\Lambda\sigma^T+\beta A^T)^T
G\hat{y}
-2q^T(\Lambda\sigma^T+\beta(t)A^T)(\Sigma\Sigma^T)^{-1}A\hat{y};
\end{eqnarray*}

3). the term that is quadratic in $\hat{y}$
\begin{eqnarray*}
&&\hat{y}^T\left(\frac{d G}{d t}+GD-A^T
(\Sigma\Sigma^T)^{-1}A-2A^T(\Sigma\Sigma^T)^{-1}(\Lambda\sigma^T+\beta
A^T)^T G\right)\hat{y}.
\end{eqnarray*}

So, if the $p, q, G$ satisfy the following three equations,
respectively,
\begin{eqnarray}\label{three-equation-1-0-0}
&&\left\{\begin{array}{l}\frac{d p}{d
t}+q^T\left(\Lambda\sigma^T+\beta A^T
\right)(\Sigma\Sigma^T)^{-1}\left(\Lambda\sigma^T+\beta A^T
\right)^T q+2r\\
-\left[2(a^T-r\mathbf{1})(\Sigma^2)^{-1}(\Lambda\sigma^T+\beta
A^T)^T-d^T \right]q
-(a^T-r\mathbf{1})(\Sigma\Sigma^T)^{-1}(a^T-r\mathbf{1})^T \\
+Tr\left[\left(\Lambda\sigma^T+\beta A^T
\right)(\Sigma\Sigma^T)^{-1}\left(\Lambda\sigma^T+\beta A^T
\right)^T G\right]=0,\\
p(T)=0,\end{array}\right. \\\label{three-equation-2-0-0-0}
&&\left\{\begin{array}{l}\frac{d q}{d t}^T+\left[q^T D+ d^T
G\right]-2(a^T-r\mathbf{1})(\Sigma\Sigma^T)^{-1}A\\
-2(a^T-r\mathbf{1})(\Sigma\Sigma^T)^{-1}(\Lambda\sigma^T+\beta
A^T)^T
G-2q^T(\Lambda\sigma^T+\beta A^T)(\Sigma\Sigma^T)^{-1}A=0,\\
q(T)=0,\end{array}\right.\\
&&\label{three-equation-3-0-0-0} \left\{\begin{array}{l}\frac{d G}{d
t}-A^T
(\Sigma\Sigma^T)^{-1}A-\left[A^T(\Sigma\Sigma^T)^{-1}(\Lambda\sigma^T+\beta
A^T)^T-\frac{1}{2}D^T\right]G\\-G\left[(\Lambda\sigma^T+\beta
A^T)(\Sigma\Sigma^T)^{-1}A-\frac{1}{2}D\right]=0,\\
G(T)=0,\end{array}\right.
\end{eqnarray}
we can determine the function $f$. Firstly, we need to claim that the second equation of (\ref{filtering}), (\ref{three-equation-1-0-0}), (\ref{three-equation-2-0-0-0}) and (\ref{three-equation-3-0-0-0}) have unique solution, respectively. In fact, it is known that the second equation of (\ref{filtering}) has a unique nonnegative definite solution; see, for example, Theorem 10.3 of \cite{Shiryaev}.
While for (\ref{three-equation-3-0-0-0}), (\ref{three-equation-2-0-0-0}) and (\ref{three-equation-1-0-0}), they are linear in $G,q,p$, respectively; thus, the solutions uniquely exist. This means that $f$ given in (\ref{f})
exactly solves (\ref{HJB-5}). Furthermore, by analysis in above subsection,  we can conclude that $H$ defined in (\ref{H}) solves (\ref{HJB-3}).
Notice that
$$
J_{XX}(t)=H_{zz}(t)=f(t,\hat{y})=\exp\left\{p(t)+q(t)^T\hat{y}+\hat{y}^T
G(t)\hat{y}\right\}>0.
$$
Thus, $H$ defined in (\ref{H}) satisfies HJB equation (\ref{HJB}). Clearly, (\ref{optimal control-1}) is equal to  (\ref{optimal control-last one0-0-0}). In the end, we need only to confirm that (\ref{optimal control-last one0-0-0}) is admissible. By classic filtering theory,
$\mathcal{G}_t$ is equal to the $\sigma$-algebra generated by
innovation process $\{v_s,s\leq t\}$ (see for example \cite{Kallianpur}.). Clearly, we have (\ref{optimal control-last one0-0-0}) is $\sigma(v_s,s\leq t)$-adapted, and thus it is admissible.
Therefore, (\ref{optimal control-last
one0-0-0}) is the optimal strategy, which make the
$\frac{1}{2}E\left[X(T)-\alpha\right]^2$ minimal.
This completes the proof.
\hfill$\square$

To this end, we give a brief discussion about the solvability in theory of (\ref{filtering})(\ref{three-equation-2-0-0})(\ref{three-equation-3-0-0}).
Clearly, $\beta(t)$ satisfies
\begin{eqnarray*}\label{convariance}
\dot{\beta}=\left(D-\Lambda\sigma^T
\Gamma^{-1} A \right)\beta+\beta(t)\left(D^T-A^T
\Gamma^{-1}\sigma\Lambda^T
\right)-\beta A^T
\Gamma^{-1}A\beta+\Lambda\Lambda^T-\Lambda\sigma^T\Gamma^{-1}
\sigma\Lambda^T
\end{eqnarray*}
with $\beta(0)=0$. Let $\overline{\beta}(s)=\beta(t),s=T-t$, then it follows
\begin{eqnarray*}
-\frac{d\overline{\beta}}{ds}=\left(D-\Lambda\sigma^T
\Gamma^{-1} A
\right)\overline{\beta}+\overline{\beta}\left(D^T-A^T
\Gamma^{-1}\sigma\Lambda^T
\right)-\overline{\beta}A^T
\Gamma^{-1}A\overline{\beta}+\Lambda\Lambda^T-\Lambda\sigma^T\Gamma^{-1}
\sigma\Lambda^T
\end{eqnarray*}
with $\bar{\beta}(T)=0$.
By known result (see for example Anderson and Moore
(1971)), $\overline{\beta}(s)$ can be represented as
\begin{eqnarray*}
\overline{\beta}(s)=K(s)L^{-1}(s),
\end{eqnarray*}
where $K(s),L(s)$ are defined as
\begin{eqnarray*}
\left[\begin{array}{c} \dot{K}\\\dot{L} \end{array}\right]=
\left[\begin{array}{cc}D^T-A^T \Gamma^{-1}\sigma\Lambda^T&-A^T
\Gamma^{-1}A\\-\Lambda\Lambda^T+\Lambda\sigma^T\Gamma^{-1}
\sigma\Lambda^T&-D+\Lambda\sigma^T \Gamma^{-1} A
\end{array}\right]\left[\begin{array}{c} K\\L
\end{array}\right],~~
\left[\begin{array}{c} K(T)\\L(T)
\end{array}\right]=\left[\begin{array}{c} I\\ 0
\end{array}\right].
\end{eqnarray*}
Therefore,
\begin{eqnarray*}
\beta(t)=\overline{\beta}(s)=K(T-t)L^{-1}(T-t).
\end{eqnarray*}
Clearly, (\ref{three-equation-3-0-0}) is a Lyapunov differential equation, which is solved by introducing the following operator
\begin{eqnarray*}
\mbox{Vec}(G(t))=\left(G(t)^{(1)T},G(t)^{(2)T},...,G(t)^{(n)T}\right)^T\triangleq
\mathbb{G}(t)\in \mathbb{R}^{n^2},
\end{eqnarray*}
where $G(t)^{(i)T}$ is the transpose of $i$-th column of of $G$. Clearly,
\begin{eqnarray*}\label{G-1}
\dot{\mathbb{G}}(t)=\mathbb{P}^{G_1}(t)\mathbb{G}(t)+\mathbb{P}^{G_2}(t),~~\mathbb{G}(T)=0,
\end{eqnarray*}
where
\begin{eqnarray*}
\mathbb{P}^G(t)&=&I\otimes\left[(\Lambda\sigma^T+\beta(t)
A^T)(\Sigma(t)\Sigma^T(t))^{-1}A-\frac{1}{2}D\right]^T\\
&&+\left[(\Lambda\sigma^T(t)+\beta(t)
A^T)(\Sigma(t)\Sigma^T(t))^{-1}A-\frac{1}{2}D\right]\otimes I,\\
\mathbb{P}^{G_2}(t)&=&\mbox{Vec}\left(A^T (\Sigma^2)^{-1}A\right).
\end{eqnarray*}
Let $\overline{\mathbb{G}}(s)=\mathbb{G}_{T-s}$.
Then
\begin{eqnarray}\label{G}
\frac{d\overline{\mathbb{G}}}{ds}=-\mathbb{P}^{G_1}(T-s)\overline{\mathbb{G}}(s)-\mathbb{P}^{G_2}(T-s),\overline{\mathbb{G}}_0=0,
\end{eqnarray}
Therefore,
\begin{eqnarray*}
\mathbb{G}(t)=\overline{\mathbb{G}}(s)=-\int_0^s\Phi(s,\tau)\mathbb{P}^{G_2}(T-\tau)d\tau,
\end{eqnarray*}
where $\Phi(\cdot,\cdot)$ is the fundamental matrix of (\ref{G}). Thus $G(t)=\mbox{Vec}^{-1}(\mathbb{G}(t))$. At last, (\ref{three-equation-2-0-0}) and (\ref{three-equation-1-0-0}) can be easily solved by the linearity of the equations.

\section{Efficient Frontier}

In this section, we proceed to derive the efficient frontier for the
original portfolio selection problem under partial information.
To begin with, we prove a lemma which shows the feasibility of the
original problem.

\begin{lem}\label{lemma}
Problem (\ref{equ007}) is feasible, and the minimal
mean-variance of the terminal wealth process is finite.
\end{lem}

\emph{{Proof.}}  The proof follows directly from results of Section 5
in \cite{Xiong-Zhou}. In the language of \cite{Xiong-Zhou},
(\ref{separation}) can be rewritten as
\begin{eqnarray}\label{bsde}
\left\{\begin{array}{l}dX(t)=[r(t)X(t)+\overline{B}(t)\Sigma^T
Z(t)]dt+Z(t)dv_t,\\
X(T)=v
\end{array}\right.
\end{eqnarray}
where $v$ is defined by Theorem 5.4 in \cite{Xiong-Zhou}
satisfying $Ev=\bar{x}$, and
$$Z(t)=(\Sigma^T(t))^{-1}\pi(t).$$
Clearly, $\mathcal{G}_t$ is equivalent to the $\sigma$-algebra generated by
innovation process $\{v_u,u\leq t\}$. By general BSDEs
theory, (\ref{bsde}) has a unique $\mathcal{G}_t$-adapted, square
integrate solution $(X(\cdot),Z(\cdot))$. Therefore, problem (\ref{equ007})
is feasible because $\Sigma^T(x)Z(t)$  is a feasible strategy. On the other hand,
by theorem 5.6 of \cite{Xiong-Zhou}, we know that the minimal
mean-variance at the terminal time point is finite. \hfill$\square$

Now, we state our main theorem.

\begin{thm}
The efficient strategy of Problem (\ref{equ007})
with the terminal expected wealth constraint $EX(T)=\bar{x}$ is
given by
\begin{eqnarray}\label{optimal control-last one-0}
\pi(t)&=&-(\Sigma(t)\Sigma^T(t))^{-1}\left[a^T-r(t)\mathbf{1}+(\Lambda\sigma^T(t)+\beta(t)A^T)^T
q(t)\right.\nonumber\\
&&~~\left.+\left(A+(\Lambda\sigma^T(t)+\beta(t)A^T)^T
G(t)\right)\hat{y}(t) \right]\left(X(t)-(\bar{x}+\gamma^*)
e^{-\int_t^Tr(s)ds}\right).\end{eqnarray} Here, $\beta(t),q(t),G(t)$
solve equation (\ref{filtering})(\ref{three-equation-2-0-0})(\ref{three-equation-3-0-0}), respectively, and $\gamma^*$ is given
by $$\gamma^*=\frac{\bar{x}-x_0e^{\int_0^Tr(s)ds}}{e^{2\int_0^Tr(s)ds}E\left[e^{2\xi_T}\right]-1}E\left[e^{2\xi_T}\right],$$
where $\xi_T$ is given by
\begin{eqnarray}\label{xi}
\xi_T&=&\int^T_0\left[r(s)-\overline{B}(s)(\Sigma\Sigma^T)^{-1}\left(V(s)+U(s)\hat{y}(s)
\right)\right]ds+\int^T_0 \left[V(s)+U(s)\hat{y}(s)
\right]^T\Sigma^{-1}dv(s)\nonumber \\&&-\frac{1}{2}\int^T_0
\left|\left[V(s)+U(s)\hat{y}(s) \right]^T\Sigma^{-1}\right|^2ds,
\end{eqnarray}
and $V(t)=a^T-r(t)\mathbf{1}+(\Lambda\sigma^T(t)+\beta(t)A^T)^T
q(t)$, $U(t)=A+(\Lambda\sigma^T(t)+\beta(t)A^T)^T G(t)$. Moreover,
the efficient frontier is given by
\begin{eqnarray}
\frac{1}{2}\left[x_0-(\bar{x}+\gamma^*)e^{-\int_0^Tr(s)ds}\right]^2E\left[e^{2\xi_T}\right]-\frac{1}{2}
\gamma^{*2}.
\end{eqnarray}
\end{thm}

\emph{{Proof}}. By Lemma \ref{lemma}, we know that the constraint
Problem (\ref{equ007}) is feasible, and its minimal
terminal mean-variance $J^*$ is finite. This means that
\begin{eqnarray}\label{optimal cost}
J^*=\max_{\gamma\in R}\min_{\pi\in
\Pi}\left\{\frac{1}{2}E[X(T)-\bar{x}]^2-\gamma
[EX(T)-\bar{x}]\right\}<\infty
\end{eqnarray}
where the equality is true by general convex constraint optimization
theory (See, for example, \cite{Luenberger}.).
By Theorem \ref{thm-1}, the
wealth equation (\ref{separation}) evolves as
\begin{eqnarray*}
dX(t)&=&\left\{r(t)X(t)-\overline{B}(t)(\Sigma(t)\Sigma^T(t))^{-1}\left[V(t)+U(t)\hat{y}(t)
\right]\left(X(t)-\alpha
e^{-\int_t^Tr(s)ds}\right)\right\}dt\\
&&-\left(X(t)-\alpha
e^{-\int_t^Tr(s)ds}\right)\left[V(t)+U(t)\hat{y}(t)
\right]^T(\Sigma(t)\Sigma^T(t))^{-1}\Sigma(t) dv(t).\end{eqnarray*}
In terms of
$z$, this equation is
\begin{eqnarray*}
\left\{\begin{array}{l}dz(t)=
\left\{r(t)-\overline{B}(t)(\Sigma\Sigma^T)^{-1}\left[V(t)+U(t)\hat{y}(t)
\right]\right\}z(t)dt\\
~~~~~~~~~~-z(t)\left[V(t)+U(t)\hat{y}(t)
\right]^T\Sigma^{-1}dv(t),\\
z(0)=x_0-\alpha e^{-\int_0^Tr(s)ds}\triangleq z_0.
\end{array}\right.\end{eqnarray*}
Clearly,
\begin{eqnarray*}
&&z(t)=z_0\exp\left\{\int^t_0\left[r(s)-\overline{B}(s)(\Sigma\Sigma^T)^{-1}\left(V(s)+U(s)\hat{y}(s)
\right)\right]dt\right.\nonumber\\
&&\left.~~~~~~~~~~~~~~~~~+\int^t_0 \left[V(s)+U(s)\hat{y}(s)
\right]^T\Sigma^{-1}dv(s)-\frac{1}{2}\int^t_0
\left|\left[V(s)+U(s)\hat{y}(s) \right]^T\Sigma^{-1}\right|^2ds
\right\}.
\end{eqnarray*}
Thus
\begin{eqnarray*}
E[z(T)]^2=z^2_0E\left[e^{2\xi_T}\right],
\end{eqnarray*}
where $\xi_T$ is defined in (\ref{xi}).
Notice that
\begin{eqnarray*}
\frac{1}{2}E[z(T)]^2=\frac{1}{2}E[X(T)-\bar{x}]^2-\gamma
[EX(T)-\bar{x}]+\frac{1}{2} \gamma^2.
\end{eqnarray*}
For any fixed $\gamma$,
\begin{eqnarray}\label{lagrange }
&&\min_{\pi\in \Pi}\left\{\frac{1}{2}E[X(T)-\bar{x}]^2-\gamma
[EX(T)-\bar{x}]\right\}\nonumber\\
&&=\frac{1}{2}\left[x_0-(\bar{x}+\gamma)e^{-\int_0^Tr(s)ds}\right]^2E\left[e^{2\xi_T}\right]-\frac{1}{2}
\gamma^2\triangleq \mathbb{J}(\gamma).
\end{eqnarray}
To obtain the optimal mean-variance value and the optimal portfolio
strategy of Problem (\ref{equ007}), we should maximize
(\ref{lagrange }) over $\gamma$ within $R$, and the finiteness is ensured
by (\ref{optimal cost}). We easily show that
(\ref{lagrange }) attains its maximum value
$\mathbb{J}\left(\gamma^*\right)$ at
\begin{eqnarray}
\gamma^*=\frac{\bar{x}-x_0e^{\int_0^Tr(s)ds}}{e^{2\int_0^Tr(s)ds}E\left[e^{2\xi_T}\right]-1}E\left[e^{2\xi_T}\right].
\end{eqnarray}
And we can assert that
$$e^{2\int_0^Tr(s)ds}E\left[e^{2\xi_T}\right]-1\neq 0.$$
If this is not true, the optimal cost will be infinite, which contradicts
(\ref{optimal cost}).  \hfill$\square$

\end{document}